\documentstyle{article}
\begin{document}

\begin{center}
{\Large \bf One dimensional potentials in $q$ space}
\end{center}

\begin{center}
\textbf{S. A. Alavi}
\end{center}

\begin{center}
\textit{High Energy Physics Division, Department of Physics, University of Helsinki\\
and\\
Helsinki Institute of Physics, FIN-00014 Helsinki, Finland.\\
On leave of absence from : Department of Physics, Ferdowsi University of Mashhad, Mashhad, P. O. Box 1436, Iran\\
E-mail: $Ali.Alavi@Helsinki.fi$\\
$ali_{-}asgharalavi@hotmail.com$}
\end{center}

\textbf{abstract.} We study the one dimensional potentials in $q$ space and the\\ new features that arise. In particular we show that the probability of tunneling of a particle through a barrier or potential step is less than the one of the same particle with the same energy in ordinary space which is somehow unexpected. We also show that the tunneling time for a particle in $q$ space is less than the one of the same particle in ordinary space.

\textbf{Pacs number.} 03.65.-w, 02.40.Gh\\

\textbf {1 Introduction}.\\
Quantum groups are a generalization of the concept of symmetries[1-5]. The mathematical theory of quantum groups arose as an abstraction
from constructions developed in the frame of the inverse
scattering method of solution of quantum integrable models, but
because of its rich and powerful structure, it has been applied
to different problems far beyond the original area. quantum groups
act on noncommutative spaces. If the space structure at short
distances (much smaller than $10^{-18}$ cm, according to the present test of quantum electrodynamics, the usual Heisenberg's commutation relations are correct at least down to $10^{-18}$ cm) shows a noncommutative property and thus governed by
quantum group symmetry, then quantum mechanics based on q-deformed              Heisenberg algebra is a possible candidate for quantum theory to study the phenomena at short distances .
Different frameworks of $q-$deformed Heisenberg algebra have been established [6-22], but physically, the one presented in Refs.[13, 17] is more clear : its relation to the corresponding $q-$deformed boson commutation relations and the limiting process of the $q-$deformed harmonic oscillator to the undeformed one are clear. By the results of string theory arguments, some applications of quantum mechanics on a non-commutative plane has been studied in [27]. Perturbative aspects of the schroedinger equation in $q$ space has been studied in [23]. There are two perturbative Hamiltonians corresponding to two ways of realizing the $q-$deformed Heisenberg algebra by the undeformed variables(which are related by the cononical transformation): One includes it in the kinetic energy term, and the other includes it in the potential.
 At the level of operators, these two perturbative Hamiltonians are different, but according to the equivalent theorem which has been demonstrated in [24]: For any regular potential which is singularity free the expectation values of these two perturbative Hamiltonian in the eigenstates of undeformed Hamiltonian are equall. This theorem establishes a reliable foundation of the perturbative calculations in $q-$deformed dynamics.\\

\textbf {2 Background of Schroedinger equation in $q$ space.}\\
The $q-$deformed phase space
variables- the position operator $X$ and the momentum operator $P$,
satisfy the following q-deformed Heisenberg algebra [13, 17]:
\begin{equation}
q^{\frac{1}{2}}XP-q^{-\frac{1}{2}}PX=iU,\hspace{.8 cm} UX=q^{-1}XU,\hspace{.8 cm} UP=qPU ,
\end{equation}
where:
\begin{equation}
X^{\dag}=X,\hspace{1. cm} P^{\dag}=P,\hspace{1. cm} U^{\dag}=U^{-1} .
\end{equation}
$U$ called the scaling operator. and satisfies the following
q-relations also [17]:
\begin{equation}
U^{-1}\equiv q^{\frac{1}{2}}[1+(q-1)X]\partial_{X},\
\overline{\partial}_{X}\equiv -q^{-\frac{1}{2}}U\partial_{X},\
P\equiv -\frac{i}{2}(\partial_{X}-\overline{\partial}_{X}),
\end{equation}
where $\overline{\partial}_{X}$ is the conjugate of
$\partial_{X}$. We observe that the operator U has been used in
the definition of the hermitian momentum, and therefore it closely
relates to properties of the dynamics and has an important role in
quantum mechanics in $q$ space. The q-deformed phase space variables
$X$, $P$ and the scaling operator $U$ can be realized in terms of
the undeformed variables $\hat{x}$ and $\hat{p}$ of the
quantum mechanics in nondeformed space [17]:
\begin{equation}
X=\frac{[\hat{z}+\frac{1}{2}]}{\hat{z}+\frac{1}{2}} \hat{x},\hspace{.5cm}
 P=\hat{p},\hspace{.5 cm} U=q^{\hat{z}},
\end{equation}
where: $\hat{z}=\frac{-i}{2}(\hat{x} \hat{p}+\hat{p} \hat{x})$ and
therefore $\hat{z}+\frac{1}{2}=-i \hat{x} \hat{p}$. $[S]$ denotes the
well known q-bracket:
\begin{equation}
[S]=\frac{q^{S}-q^{-S}}{q-q^{-1}} .
\end{equation}
From (4), one can show that $X$ can be represented as a function of $x$ and
$p$ as follows:
\begin{equation}
X=i(q-q^{-1})^{-1}(q^{(\hat{z}+\frac{1}{2})}-q^{-(\hat{z}+\frac{1}{2})})
\hat{p}^{-1} .
\end{equation}
If we let $q=e^{\theta}=1+\theta$, with $0<\theta\ll 1$, the
perturbative expansion of $X$ to the order $\theta^{2}$, is given
by [23]:
\begin{equation}
X=\hat{x}+\theta^{2} g(\hat{x},\hat{p}), \hspace{1.8 cm} g(\hat{x},\hat{p})=
-\frac{1}{6}(1+\hat{x}\hat{p}\hat{x}\hat{p})\hat{x}.
\end{equation}
The Hamiltonian is :
\begin{equation}
H(X,P)=\frac{1}{2m}P^{2}+V(X).
\end{equation}
The singularity free potential $V(X)$, can be expressed in terms
of the undeformed variables $x$ and $p$ as [23] :
\begin{equation}
V(X)=V(\hat{x})+\theta^{2}\hat{H}_{I}(\hat{x},\hat{p}),
\end{equation}
where
\begin{equation}
\hat{H}_{I}(\hat{x},\hat{p})=\sum_{k=1}^{\infty}\frac{V^{(k)}(0)}{k!}(\sum_{i=0}^{k-1}
\hat{x}^{(k-1)-i} g(\hat{x},\hat{p}) \hat{x}^{i}),
\end{equation}
 where $V^{k}(0)$ is the k-th derivative of $V(x)$ at $x=0$(x is the spectrum of $\hat{x}$). Then we can write:
\begin{equation}
H(X,P)=\frac{1}{2m}\hat{p}^{2}+V(\hat{x})+\theta^{2}H_{I}=H_{0}+\theta^{2}H_{I},
\end{equation}
where $H_{0}=\frac{1}{2m}\hat{p}^{2}+V(\hat{x})$ is the Hamiltonian in ordinary space which its eigenvalues and eigenfunctions are known :
\begin{equation}
H_{0}\phi_{n}=E^{0}_{n}\phi_{n}.
\end{equation}
We search for the eigenvalues and eigenfunctions of the Hamiltonian $H$ :
\begin{equation}
H\psi_{n}=(H_{0}+\theta^{2}H_{I})\psi_{n}=E_{n}\psi_{n},
\end{equation}
where $E_{n}=E_{n}^{0}+\Delta E_{n}^{\theta}$. The $\phi_{i}$ are eigenfunctions of the Hamiltonian $H_{0}$, and form a complete set therefore we can expand $\psi_{n}$ in a series as follows :
\begin{equation}
\psi_{n}=\phi_{n}+\sum_{m\neq n}C_{nm}(\theta^{2})\phi_{m} ,
\end{equation}
where $C_{nk}(0)=0$. We can also expand $C_{nk}(\theta^{2})$ and $E_{n}$ as :
\begin{equation}
C_{nm}(\theta^{2})=\theta^{2}C_{nm}^{(1)}+\theta^{4}C_{nm}^{(2)}+...\hspace{.5 cm .}
\end{equation}
\begin{equation}
E_{n}=E_{n}^{0}+\theta^{2}\Delta E_{n}^{(1)}+\theta^{4}\Delta E_{n}^{(2)}+...\hspace{.5 cm .}
\end{equation}
Then the Schroedinger equation takes the form :\\

$(H_{0}+\theta^{2}H_{I})(\phi_{n}+\sum_{m\neq n}\theta^{2}C_{nm}^{(1)}\phi_{m}+\sum_{m\neq n}\theta^{4}C_{nm}^{(2)}\phi_{m}+...)=$
\begin{equation}
(E^{0}_{n}+\theta^{2}\Delta E_{n}^{(1)}+\theta^{4}\Delta E_{n}^{(2)}+...)(\phi_{n}+\sum_{m\neq n}\theta^{2}C_{nm}^{(1)}\phi_{m}+\sum_{m\neq n}\theta^{4}C_{nm}^{(2)}\phi_{m}+...).
\end{equation}
Using uniqueness theorem we will have a series of equations. The first one is :
\begin{equation}
H_{0}\sum_{m\neq n}C^{(1)}_{nm}\phi_{m}+H_{I}\phi_{n}=E_{n}^{0}\sum_{m\neq n}C^{(1)}_{nm}\phi_{m}+\Delta E_{n}^{(1)}\phi_{n} .
\end{equation}
Using $H_{0}\phi_{m}=E_{m}^{0}\phi_{m}$, this leads to :
\begin{equation}
\Delta E_{n}^{(1)}\phi_{n}=H_{I}\phi_{n}+\sum_{m\neq n}(E_{m}^{0}-E_{n}^{0})C_{nm}^{(1)}\phi_{m} .
\end{equation}
Taking the scalar product with $\phi_{n}$ and using the orthogonality condition for $\phi_{i}$, we will have :
\begin{equation}
\theta^{2}\Delta E_{n}^{(1)}=<\phi_{n}\mid\theta^{2}H_{I}\mid\phi_{n}> .
\end{equation}
If we take the scalar product with $\phi_{m}$, for $m\neq n$, we obtain :
\begin{equation}
\theta^{2}C_{nm}^{(1)}=\frac{<\phi_{m}\mid\theta^{2}H_{I}\mid\phi_{n}>}{E_{n}^{0}-E_{m}^{0}} .
\end{equation}
These are two important formulas. The value of $\theta^{2}\Delta E_{n}^{(1)}$ has been calculated in [23].
\begin{equation}
\theta^{2}\Delta E_{n}^{(1)}=\frac{\theta^{2}}{6}\int_{-\infty}^{\infty} dx
\psi_{n}^{(0)*}(x) (V(x)\{1-4m
x^{2}[V(x)-E_{n}]\}-\frac{2}{3}m E_{n}x^{3}(\frac{d
V(x)}{dx})) \psi_{n}^{(0)}(x),
\end{equation}
or :
\begin{equation}
\theta^{2}\Delta E_{n}^{(1)}=\frac{\theta^{2}}{6}\int_{-\infty}^{\infty} dx
\psi_{n}^{(0)*}(x) [V(x)-E_{n}]\{1-4m
x^{2}[V(x)-E_{n}]\} \psi_{n}^{(0)}(x),
\end{equation}
where $\psi_{n}^{(0)}(x)$ and $E_{n}$ are unperturbed wave function
and energy respectively. The important point is, since $<n\mid V\mid n> < E$, we see from equ.(23) that, the energy shift is always negative :
\begin{equation}
\theta^{2}\Delta E_{n}^{(1)} <0
\end{equation}
By use of equs.(14) and (21), one can obtain    the wave functions in q space.\\

\textbf {3 One dimensional potentials in $q$ space}.\\
In this section we study some important one dimensional potentials in $q$ space. They are of interest because they may show some new features and improve our knoweldge about $q$ space and because many physical situation can be considered as one dimensional even though the real world is three dimensional. we start with the potential step.\\

\textbf {3.1 potential step}.\\
Consider a particle which is moving in the positive x-direction with energy $E$ and passes  a potential step at $x=0$ :\\
$V(x)=0$  \hspace{1. cm}      $x<0$\\
$V(x)=V_{0}$\hspace{1. cm}          $x>0$\\
We assume $E_{0}< V_{0}$. The most general solution of the Schroedinger equation for $x<0$ is:\\
\begin{equation}
u(x)= e^{ikx} +Re^{-ikx}.
\end{equation}
where $k=(2mE_{0})^{\frac{1}{2}}$,(we assume $\hbar =1$). For $x>0$, we have only the transmitted wave:
\begin{equation}
u(x)= Te^{-Qx}
\end{equation}
where $Q=[2m(V_{0}-(E_{0}+\theta^{2}\Delta E^{(1)}))]^\frac{1}{2}$. To the first order of perturbation, $\theta^{2}\Delta E^{(1)}$ is vanish and to the second order it is given by :
\begin{equation}
\theta^{2}\Delta E^{(1)}=\frac{E_{0}\theta^{2}}{3(2m(V_{0}-E_{0}))^{\frac{1}{2}}}(1-\lambda (\theta)),
\end{equation}
 where $\lambda (\theta)=\frac{1+\frac{1}{6}(V_{0}-E_{0})^{2}\theta^{2}}{1+\frac{1}{6}\theta^{2}}>1$. We observe that the energy shift $\theta^{2}\Delta E^{(1)}$, is negative, and therefore the value of $Q$ is larger than its counterpart in the nondeformed case $\beta$, $\beta=[2m(V_{0}-E_{0})]^\frac{1}{2}$. This means that the exponential in the wave function for $x>0$ damps more rapidly than the nondeformed case, and therefore the probability of the penetration of the particle in to the forbidden region in $q$ space is less than the one of the particle with the same energy in nondeformed case.
One can calculate the reflected or the transmitted coefficients using the continuity of the wave function and its derivative at $x=0$:
\begin{equation}
T^{2}_{\theta}=\frac{4E_{0}}{V_{0}+\frac{E_{0}\theta^{2}}{3(2m(V_{0}-E_{0}))^{\frac{1}{2}}}(\lambda (\theta)-1)}.
\end{equation}
Comparing to the expression for $T$ in nondeformed case $T^{2}_{\theta=0}=\frac{4E_{0}}{V_{0}}$, we observe that the denominator of $T^{2}_{\theta}$ is larger than the denominator of $T^{2}_{\theta=0}$, and therefore we have $T^{2}_{\theta}<T^{2}_{\theta=0}$.\\

\textbf{3.2 The potential well}.\\
We consider the following potential well :\\
$V(x)=0 \hspace{1.5 cm}        x<-a$ .\\
$V(x)=-V_{0} \hspace{1. cm}   -a<x<a$ .\\
$V(x)=0  \hspace{1.5 cm}        a<x$ .\\
We first study the case $E_{0}>0$. The solutions of the Schroedinger equation in nondeformed case are [25]
\begin{equation}
u(x)= e^{ikx}+R e^{-ikx},\hspace{1. cm} x<-a ,
\end{equation}
\begin{equation}
u(x)= A e^{i\beta x}+B e^{-i\beta x},\hspace{1. cm} -a<x<a ,
\end{equation}
\begin{equation}
u(x)= T e^{ikx},\hspace{1.3 cm} x>a ,
\end{equation}
where: $k^{2}=2mE_{0}$ and $\beta^{2}=2m(E_{0}+V_{0})$. For the special case $\sin(2\beta a)=0$, there is no reflection [25] and this is a model of scattering of low energy electrons(.1 ev) by a noble  gas atoms for example, neon and argon. For studing the q analouge of this situation we should calculate the energy shift and the $q$ wave functions :\\

$\theta^{2}\Delta E_{n}^{(1)}=$
\begin{equation}
\frac{\theta^{2}V_{0}}{6}\{(2-\frac{16ma^{2}V_{0}}{3})a-\frac{4ma^{3}V_{0}}{3}+(\frac{1}{2}-a+\frac{8ma^{3}V_{0}}{3})\frac{8mV_{0}a^{2}}{n^{2}\pi^{2}}+\frac{n^{2}\pi^{2}a}{3}-\frac{64m^{2}a^{2}V_{0}}{n^{4}\pi^{4}}\}.
\end{equation}
One can obtain the wave functions in $q$ space, using equs.(14) and (21). The $\theta^{2}C^{(1)}_{nm}$ coefficients are as follows :\\

$\theta^{2} C^{(1)}_{nm}=-\frac{2}{3}\frac{mV_{0}(V_{0}-E_{n}^{0})\theta^{2}}{E^{0}_{n}-E^{0}_{m}}\{(A^{2}+B^{2})[\frac{2}{(\beta_{m}-\beta_{n})}a^{2}\sin((\beta_{m}-\beta_{n})a)+\frac{4}{(\beta_{m}-\beta{n})^{2}}$\\

$\cos((\beta_{m}-\beta_{n})a)-\frac{4}{(\beta_{m}-\beta_{n})^{3}}\sin((\beta_{m}-\beta_{n})a)]+(A^{*}B+B^{*}A)[\frac{2}{(\beta_{n}+\beta_{m})}a^{2}$\\
\begin{equation}
\sin((\beta_{n}+\beta_{m})a)+\frac{4}{(\beta_{n}+\beta_{m})^{2}}\cos((\beta_{n}+\beta_{m})a)-\frac{4}{(\beta_{n}+\beta_{m})^{3}}\sin((\beta_{n}+\beta_{m})a)]\}
\end{equation}

where $\beta_{n}^{2}=\frac{n^{2}\pi^{2}}{4a^{2}}$ and $E^{0}_{n}=-V_{0}+\frac{n^{2}\pi^{2}}{8ma^{2}}$. In $q$ space the particle has the\\

energy $E_{n}^{0}+\theta^{2}\Delta E_{n}^{(1)}$. The coefficients $A$ and $B$ are as follows :\\

$\hspace{.2 cm}A=\frac{\beta+k}{2\beta}\cos((\beta-k)a)+\frac{\beta-k}{2\beta N}(L\cos((k+\beta)a)-M\sin((\beta+k)a))$
\begin{equation}
+i\{\frac{\beta+k}{2\beta}\sin((\beta-k)a)+\frac{\beta-k}{2\beta N}(L\sin((\beta+k)a)+M\cos((\beta+k)a)\},
\end{equation}
$\hspace{.5 cm}B=\frac{\beta-k}{2\beta}\cos((\beta+k)a)+\frac{\beta+k}{2\beta N}(L\cos((k-\beta)a)-M\sin((k-\beta)a))$
\begin{equation}
+i\{\frac{k-\beta}{2\beta}\sin((\beta+k)a)+\frac{\beta+k}{2\beta N}(L\sin((k-\beta)a)+M\cos((k-\beta)a)\},
\end{equation}
where :\\

$\hspace{.25 cm}L=(k^{4}-\beta^{4})\sin^{2}(2\beta a)\cos(2ka)+k\beta(\beta^{2}-k^{2})\sin(2ka)\sin(4\beta a).$\\

$M=(\beta^{4}-k^{4})\sin^{2}(2\beta a)\sin(2ka)+k\beta(\beta^{2}-k^{2})\cos(2ka)\sin(4\beta a).$\\

$N=(2k\beta \cos(2\beta a))^{2}+(\beta^{2}+k^{2})^{2}\sin^{2}(2\beta a).$\\

Now we study the case $E<0$, the solutions of the Schroedinger equation in ordinary space for outside the well are as follows [25]:
\begin{equation}
u(x)= c_{1} e^{kx}     \hspace{1. cm}  x<-a .
\end{equation}
\begin{equation}
u(x)= c_{2} e^{-kx}  \hspace{1. cm}    a<x .
\end{equation}
The solutions inside the well are either even in $x$ ( $\cos\beta_{n}x$)or odd in $x$ ($\sin\beta_{n}x$) with $\beta^{2}_{n}=2m(V_{0}-\mid E_{n}^{0}\mid)>0$ where $E_{n}^{0}(even)=(n+\frac{1}{2})\pi$ and $E_{n}^{0}(odd)=n\pi$. The energy shift corresponding to even and odd solutions are :
\begin{equation}
\theta^{2}\Delta E_{n}^{(1)}(even)=\frac{\theta^{2}V_{0}}{6}[1-4ma^{2}V_{0}(V_{0}-n\pi-\frac{\pi}{2})(\frac{1}{3}-\frac{1}{2(n+\frac{1}{2})^{2}\pi^{2}})] .
\end{equation}
\begin{equation}
\theta^{2}\Delta E_{n}^{(1)}(odd)=\frac{\theta^{2}V_{0}}{6}[1-4ma^{2}V_{0}(V_{0}-n\pi)(\frac{1}{3}-\frac{1}{2n^{2}\pi^{2}})] .
\end{equation}
and the coefficients $C^{(1)}_{nm}$ for even and odd solutions can be obtained from (34) by choosing $A=B=1$ and $A=-B=1$ respectively.\\

\textbf {3.3 The potential barrier}.\\
As we know, like drift and diffusion, tunneling is also a basic mechanism of carrier transport. in the following we consider the rectangular barrier :\\
$V(x)=0  \hspace{1.2 cm}   x<-a$ .\\
$V(x)=V_{0} \hspace{.8 cm} -a<x<a$ .\\
$V(x)=0  \hspace{1.2 cm}      a<x$ .\\
We study the case $E<V_{0}$. The solutions of the Schroediger equation in $q$ space are :
\begin{equation}
u(x)= e^{ikx}+ R e^{-ikx}\hspace{1. cm}   x<-a .
\end{equation}
\begin{equation}
u(x)= C e^{Qx}+ D e^{-Qx}\hspace{1.4 cm}     \mid x\mid<a .
\end{equation}
\begin{equation}
u(x)= T e^{ikx}\hspace{2.3 cm}    x>a .
\end{equation}
The energy of the particle in the region $-a<x<a$, is $V_{0}-(E_{0}+\theta^{2}\Delta E^{(1)})$ with :\\

$\theta^{2}\Delta E^{(1)}=\frac{V_{0}\theta^{2}}{6}\{(C^{2}+D^{2})\frac{\sinh(2\beta a)}{\beta}[1-4m(V_{0}-E_{0})(a^{2}+\frac{1}{2\beta^{2}})]+$
\begin{equation}
4ma(C^{2}+D^{2})(V_{0}-E_{0})\frac{\cosh(2\beta a)}{\beta^{2}}+2a(C^{*}D+D^{*}C)[1-\frac{4ma^{2}}{3}(V_{0}-E_{0}]\},
\end{equation}
where $\beta^{2}=[2m(V_{0}-E_{0})]$. The coefficients $C$ and $D$ can be obtained from $A$ and $B$ (equs.(34),(35)) by replacing $\beta\rightarrow iQ$ as follows :\\

$C=B(\beta\rightarrow iQ)\hspace{4 cm}D=A(\beta\rightarrow iQ)$.\\

Since the energy shift is always negative (see equ.(24)), we have :\\

\hspace{.95cm}$Q=\sqrt{2m[V_{0}-(E+\theta^{2}\Delta E^{(1)})]}>\sqrt{[2m(V_{0}-E)]}=\beta$ .\\
Matching wave functions and their derivatives at $x=\pm a$, give the following expression for the ratio of transmitted flux to incident flux :
\begin{equation}
\mid T_{\theta}\mid^{2}=\frac{(2Qk)^{2}}{(k^{2}+Q^{2})^{2}\sinh^{2}(2Qa)+(2Qk)^{2}} .
\end{equation}
When $Qa$ is large, $\sinh^{2}(2Qa)\approx e^{4Qa}$, and we have :
\begin{equation}
\mid T_{\theta}\mid^{2}=(\frac{2kQ}{k^{2}+Q^{2}})^{2} e^{-4Qa} .
\end{equation}
Since $Q>\beta$, the exponential in the expression for $\mid T_{\theta}\mid^{2}$, damps more rapidly than the nondeformed case:
\begin{equation}
\mid T_{\theta=0}\mid^{2}=(\frac{2k\beta}{k^{2}+\beta^{2}})^{2} e^{-4\beta a} .
\end{equation}
and we have :
\begin{equation}
\frac{\mid T_{\theta}\mid^{2}}{\mid T_{\theta=0}\mid^{2}}=(\frac{Q}{\beta}\frac{k^{2}+\beta^{2}}{k^{2}+Q^{2}})^{2}e^{-4(Q-\beta)a}<1 .
\end{equation}
Again this means that, comparison to the nondeformed case the probability of penetration of the particle in to the forbidden region is reduced.\\
So far we discussed the rectangular barriers. For studing the barriers with arbitrary shapes, physicits use WKB approximation, which leads to the following expression for $T$ [25-26]:
\begin{equation}
\mid T\mid^{2}\approx e^{-2\int^{x_{2}}_{x_{1}} dx \sqrt{2m(V(x)-E_{0})}},
\end{equation}

The integral in the exponent runs over the classicaly forbidden domain. We have the following expression for $\mid T_{\theta}\mid^{2}$ :
\begin{equation}
\mid T_{\theta}\mid^{2}\approx e^{-2\int^{x_{2}}_{x_{1}} dx \sqrt{2m((V(x)+\mid\theta^{2}\Delta E^{(1)}\mid)-E_{0})}}.
\end{equation}

This can be written as :
\begin{equation}
\mid T_{\theta}\mid^{2}\approx e^{-2\sqrt{2m}\int^{x_{2}}_{x_{1}} dx (\sqrt{V(x)-E_{0}}+\frac{1}{2}\mid\theta^{2}\Delta E^{(1)}\mid(V(x)-E_{0})^{\frac{-1}{2}})}.
\end{equation}
Since $V(x)>E_{0}$ for $x_{1}<x<x_{2}$, the second term in the exponent is always positive and therefore we have :
\begin{equation}
\mid T_{\theta}\mid^{2}<\mid T_{\theta=0}\mid^{2} .
\end{equation}
Noe we study another important quantity in tunneling process namely the tunneling time(the time which tunneling process takes) which is substantial for physical applications and for a proper understanding of quantum theory. For rectangular barrier it is given by [26]:

\begin{equation}
\tau=\frac{1}{4(V_{0}-E_{0})} .
\end{equation}
for the $q-$deformed case we have :
\begin{equation}
\tau_{\theta}=\frac{1}{4((V_{0}+\mid\theta^{2}\Delta E^{(1)}\mid)-E_{0})} .
\end{equation}
As we observe, the denominator of $\tau_{\theta}$ is larger than the denominator of $\tau$, and therefore we have :
\begin{equation}
\tau_{\theta}<\tau .
\end{equation}
which means that the tunneling time for a particle in $q$ space is less than the one of a particle with the same energy in ordinary space.\\

\textbf{Acknowledgment.}\\
I would like to thank Professor P. Pre\v{s}najder and Professor Jian-zu Zhang for their  careful reading of the manuscript and for their valuable comments. I acknowledge Professor P. Kulish for helpful discussions. I am also very grateful to Professor Masud Chaichian for his warm hospitality during my visit to the university of Helsinki. This work is partialy supported by the Ministry of Science, Research and Tecknology of Iran.\\\\

\textbf{References.}\\

1. P. Kulish and Yu. Reshtikhin, J. SoV. Math. 23 (1983) 2435 (translated from : Zapiski Nautch. Seminarov LOMI 101 (1981) 101.\\
2. L. D. Faddeev and L. A. Takhtajan, Lectures Notes in Physics 246 (1986) 166.\\
3. V. G. Drinfel'd, in Proc. Of the International congress of Mathematics (Berkley, 1986), American Mathematics Society, 1987, P. 798.\\
4. M. Jimbo Lett. Math. Phys 10 63, 1985\\
            Lett. Math. Phys 11 247, 1986.\\
5. M. Chaichian and A. Demichev, Introduction to quantum groups, World Scientific, Singapore (1996).\\
6. A. J. Macfarlane, J. Phys. A22 (1989) 4581.\\
7. L. C. Bidenharn, J. Phys. A22 (1998) L873.\\
8. M. Chachian and P. Kulish, Phys. Lett. B234 (1990) 72; P. P. Kulish and E.     V. Damashinski, J. Phys. A23 (1990) L415.\\
9. C.-P. Sun and H. Fu, J. Phys. A22 (1998) L983.\\
10. J. Schwenk and J. Wess, Phys. Lett. B291 (1992) 273.\\
11. S. Codrianski, Phys. Lett. A184 (1993) 381.\\
12. A. Kempf, J. Math. Phys. 35 (1994) 4483 [hep-th/9311147].\\
13. A. Hebecher, S. Schreckenberg, J. Schwenk, W. Weich and J. Wess, Z. Phys. C    64 (1994) 355.\\
14. R. J. McDermott and A. I. Solomon, J. Phys. A 27 (1994) L15.\\
15. C. A. Nelson and M. H. Fiels, Phys. Rev. A51 (1995) 2410.\\
16. A. Lorek and J. Wess, Z. Phys. C67 (1995)671 [q-alg/9502007].\\
17. M. Fichtmuller, A. Lorek and J. Wess, Z. Phys. C71 (1996) 533 [hep-th/9511     106].\\
18. A. Lorek, A. Ruffing and J. Wess, Z. Phys. C74(1997) 364[hep-th/9605161].\\
19. Jian-zu Zhang, Phys. Lett. B440 (1998) 66.\\
20. Jian-zu Zhang, Phys. Lett. A262 (1999) 125; P. Osland and J. Zhang, hep-th     /0010176.\\
21. B. L. Cerchiai, R. Hinterding, J. Madore and J. Wess, Eur. Phys. J. C8(199     9) 547 [math.qa/9809160].\\
22. J. Zhang, Chin. Phys. Lett. 17 (2000) 91; Phys. Lett. B477 (2000) 361.\\
23. Jian-zu Zhang and P. Osland, Eur. Phys. J. C20 (2001) 393-396 [hep-th/0102     014].\\
24. Jian-zu Zhang, Phys. Lett. B 517 (2001) 210-214 [hep-th/0108111].\\
25. Stephen Gasioriwicz, Quantum Physics, John Wiley $\&$ Sons, Inc, 1974.\\
26. D. K. Roy, Quantum Mechanical tunnelling And Its Applications, World Scientific Co Pte Ltd, 1986.\\
27. D. Kochan and M. Demetrian, hep-th/0102050.

\end{document}